\documentclass{elsart}

\usepackage{amssymb,graphicx}
\newcommand{\beqa}{\begin{eqnarray}}
\newcommand{\eeqa}{\end{eqnarray}}
\newcommand{\beq}{\begin{equation}}
\newcommand{\eeq}{\end{equation}}
\newcommand{\arl}{\begin{array}{l}}
\newcommand{\earl}{\end{array}}
\newcommand{\fract}[2]{{\textstyle\frac{#1}{#2}}}
\newcommand{\fracd}[2]{{\displaystyle\frac{#1}{#2}}}
\newcommand{\Tr}{{\mathrm Tr}}
\def\ep{\varepsilon}
\def\pa{\partial}
\def\th{\theta}

\begin{document}

\begin{frontmatter}
\title{Kaon Condensates, Nuclear Symmetry Energy and Cooling
of Neutron Stars}
\author[ifj]{S. Kubis\corauthref{cor}},
\corauth[cor]{Corresponding author.}
\ead{kubis@alf.ifj.edu.pl}
\author[ifj,ifuj]{M. Kutschera}
\address[ifj]{H. Niewodnicza\'{n}ski Institute of Nuclear Physics,
Radzikowskiego 152, 31-342 Krak\'ow, Poland}
\address[ifuj]{Institute of Physics, Jagiellonian University, 
Reymonta 4, 30-059 Krak\'ow, Poland}
\begin{abstract}

The cooling of neutron stars by URCA processes in the kaon-condensed neutron
star matter for various forms of nuclear symmetry energy is investigated. The
kaon-nucleon interactions are described by a chiral lagrangian. Nuclear matter
energy is parametrized in terms of the isoscalar contribution  and the nuclear
symmetry energy in the isovector sector. High density behaviour
of nuclear symmetry energy  plays an essential role in determining the
composition of the kaon-condensed neutron star matter which in turn affects
the cooling properties. We find that the symmetry energy which decreases at
higher densities makes the kaon-condensed neutron star matter fully
protonized.
This effect inhibits strongly direct URCA processes resulting in  slower
cooling of neutron stars as only kaon-induced URCA cycles are present. In
contrast, for increasing symmetry energy direct URCA
processes are allowed  in the almost whole density range where the kaon
condensation exists.

\end{abstract}
\begin{keyword}
dense matter \sep nuclear symmetry energy \sep 
kaon condensation \sep  neutron star cooling
\PACS 21.65.+f \sep 26.60.+c \sep 97.60.Jd 

\end{keyword}

\end{frontmatter}

\section{Introduction}

The possibility of the kaon condensation in dense nuclear matter,
proposed by Kaplan and Nelson \cite{Kaplan:1986yq},
is recently a subject of intensive theoretical   and
 experimental research \cite{Lee:1996ef,Li:1997zb}. Such a
kaon-condensed phase of dense matter is of direct astrophysical relevance
 as it would form in neutron stars and could affect their properties. In	
this paper we study how sensitive is the cooling of neutron stars to the
 presence of the kaon condensates. A particularly important problem in
 this regard, which we address
here, is to assess the influence of the uncertainty of the high density
behaviour of nuclear symmetry energy on the charged kaon condensation itself.
{Other strange hadrons, like neutral kaons and hyperons could also
appear in neutron star matter. Their inclusion here, however, would blur the
role of the nuclear symmetry energy as it would introduce additional 
uncertainty due to poorly known nucleon-hyperon interactions. 
The analysis reported here will be extended to account for hyperons in a 
forthcoming paper. 
Model calculations show that hyperons are likely to influence  the results 
concerning
kaon condensation \cite{Knorren:1995ds}.}

Among theoretical approaches we may distinguish those based on the
chiral  theory \cite{Kaplan:1986yq} and the ones using the meson
exchange picture. The former ones have their roots in the
spontaneously broken $ SU(3)\!\times\!SU(3) $
symmetry. It gives a solid base to study kaon-nucleon interactions
where kaons are treated as pseudo-Goldstone bosons.
In the meson exchange picture,
the kaon couplings  to nucleons are realized indirectly through
the exchange of vector and scalar mesons \cite{Glendenning:1999ak}. The
predictive power of chiral theory is lost in such approaches. In order to
study the kaon condensation in the neutron star matter a realistic model of
nucleon-nucleon interactions should be considered.

The nucleon-nucleon part of the interaction Hamiltonian is usually
taken from other theories of nuclear matter, such as e.g. many-body
models with {\em realistic potentials} or relativistic mean field (RMF)
theory. Realistic potential approach is firmly based
on experiment. Two-nucleon potentials are fitted to
the scattering data. Three-body potentials must be added to get
agreement with the binding energy of light nuclei ($^3$H, $^4$He) and
the saturation properties of nuclear matter. The  energy of nuclear
matter is usually derived in a variational approach  \cite{Wiringa:1988tp}
or by using other techniques as e.g. some version of the  Brueckner
theory.

The RMF approach starts from postulating the Lagrangian for nucleons
and mesons which takes into account relevant internal symmetries
(as e.g. isospin). The meson-nucleon interactions are assumed to be
of a Yukawa type with coupling constants treated as
free parameters which are fitted
to the saturation point properties obtained from the mass formula and
from breathing modes of giant resonances \cite{glend-book}.

{Unfortunately, up to now there is no consistent description of the 
ground state properties of nuclear matter in terms of the chiral theory
with its free-space parameters.
Recently some novel approaches  were proposed.
One of them starts from the chiral Lagrangian \cite{Mao:1998nv}
but still treats its crucial parameters in the spirit of RMF model -- 
fitting them to the saturation point properties. A 
different idea based on the power counting  scheme was presented in 
\cite{Kaiser:2001jx}, but its extrapolation to
higher densities seems not to be well justified.}

In the next section we discuss the high density behaviour of the nuclear
symmetry energy. One should stress that the symmetry energy governs the
chemical composition
of  nuclear matter in the beta equilibrium and hence it determines the
particle species present in  neutron stars. In Sect.\ref{sec_model}
the chiral model kaon-nucleon interactions is derived.
In Sect.\ref{sec_thermod}
the kaon condensate properties
and the role of nuclear symmetry energy in  a kaon-condensed matter are
studied. In particular we assess how uncertainty of the behaviour of
the symmetry energy at high densities influences conclusions regarding the
formation and properties of the kaon-condensed neutron star matter.
Implications  for the cooling of neutron stars are investigated in
Sect.\ref{sec_cooling}.

\section{Nuclear symmetry energy}
\label{sec_symmen}

A common feature of the RMF models is that the nuclear symmetry energy,
whose value at the saturation density is well known to be about 30 MeV,
strongly increases with density of the nuclear matter.
It is not the case for realistic potential models of nuclear matter in
the variational approach \cite{Wiringa:1988tp}, where
the symmetry energy saturates and then bends over at higher densities.
The high-density behaviour of symmetry energy is the worst known
property of dense matter \cite{Kutschera,Li:2002qx}, with different nuclear
models giving
contradictory predictions, as in the case of the RMF theory and
realistic potential models which provide incompatible
predictions of the nuclear symmetry energy at high densities.
Such divergent predictions are the source of a major uncertainty of high
density properties of neutron star matter \cite{Kutschera:2000hq},
 including the formation
of kaon condensates \cite{Kubis:1999sc}.

 To assess the influence of the symmetry energy we explore various
forms of its high density behaviour as provided by different models of
dense nuclear matter. We use three types of the high density symmetry
energy obtained from the realistic potential theory
\cite{Wiringa:1988tp} and also the symmetry energy derived from the RMF
model which are displayed in Fig.\ref{rys_Es}. Some preliminary results
are given in \cite{Kubis:1999sc}, where it was shown that low values of
the symmetry energy may exclude the condensation. However, if the
condensate is formed, the  proton fraction goes very high, even to complete
protonization of nuclear matter,
$x=1$. These results encourage us to explore the implications for
cooling of neutron stars.

The nuclear symmetry energy is defined as
\[
E_s(n) = \frac{1}{8}\frac{\pa^2\!E}{\pa x^2}(n,\fract{1}{2}),
\]
where $E(n,x)$ is the energy per baryon as a function of the baryon
number density, $n$, and the proton fraction, $x$.
The nuclear symmetry energy plays an essential role in the neutron
star matter when the beta equilibrium among all its components is
achieved. From thermodynamic identities it follows that
 $\pa E /\pa x = \mu_n - \mu_p $, where $\mu_n, \mu_p$ are neutron
 and proton chemical potentials. The isospin symmetry allows us to
expand $E(n,x)$ in even powers of $(x-\frac{1}{2})$. The expansion
up to the second order is enough, i.e. the term
$\sim (x-\frac{1}{2})^4$ seems to be negligible \cite{bombaci-lee}.
We may thus write with a sufficient accuracy
\beq
4 (1-2x) E_s = \mu_n - \mu_p .
\eeq
The beta equilibrium yields a sequence of equalities between
chemical potentials for particles present in the system and hence
the symmetry energy governs the chemical composition of
the neutron star matter. For example, in the simplest case of
$n, p, e$ matter, we have $4(1-2x) E_s = \mu_n - \mu_p = \mu_e$.
Taking into account the neutrality condition, $k_e = k_p$, and
neglecting the electron mass we obtain the proton fraction
\beq
x = \frac{1}{2} (1- \frac{k_p}{4 E_s}).
\label{x-npe}
\eeq
At the saturation density, where the value of symmetry energy is known
to be $E_s(n_0)=a_4=30$  MeV \cite{myers}, the proton abundance is about
4\%.
As the proton Fermi momentum is slowly varying with density, $k_p \sim
n_p^{1/3}$, we conclude that if $E_s$  increases faster than $n^{1/3}$
then the neutron star matter becomes more and more symmetric at high
densities: $x \rightarrow \frac{1}{2}$,
(muons, produced when $k_e > m_\mu$, make this effect even
stronger because  protons have to neutralize additional negative charge
carriers). 
{This is the case of RMF models. In its basic version where the only isovector
interaction is due to Yukawa coupling
to the $\rho$ meson, the symmetry energy takes the form \cite{Kubis:1997ew}:
\beq
E_s(n) = \frac{k_F^2}{6 \sqrt{k^2_F + m^2_*}} + \frac{g_\rho^2}{8 m^2_\rho} n ,
~~~~k_F=(\fract{3 \pi^2}{2} n)^{1/3}
\eeq
The first term is purely kinematic, coming from the energy difference of 
proton and neutron Fermi seas. The second term comes just from interaction and 
is
determined by the coupling constant. As we see, $E_s$ grows  slightly faster 
than linear
 with density and this is its typical behaviour in RMF theories\footnote{In
our earlier 
work \cite{Kubis:1997ew} we tested if the inclusion of isovector-scalar 
component due to the 
 $\delta$ meson, could change the rapid growth of the symmetry energy.}.
}
In an opposite case, when the symmetry energy decreases with
density,
one expects the neutron star matter to be
more neutron-rich, $x \rightarrow 0$, at high densities. This simple
picture need not be always true if kaons appear.
Although kaons, similarly as electrons and muons, carry negative
charge, their presence in the neutron star matter may have unexpected
consequences.

\section{Model for $K-N$ interactions}
\label{sec_model}

In our investigation we use the effective chiral theory proposed by
Kaplan and Nelson \cite{Kaplan:1986yq} and  extensively explored in
subsequent works.
The Lagrangian for kaons represents a nonlinear effective theory which
is not renormalizable, but it makes a good starting point for
{\em chiral perturbation theory}
\cite{Manohar:1984md} (we use the same symbols as in \cite{Thorsson:1994bu},
for  details see Appendix \ref{app-a1}).
\beqa
{\mathcal  L_\chi} & = & \fract{f^2}{4}\Tr \pa_\mu U \pa^\mu U^\dagger +
        \Tr\bar{B}(i \gamma^\mu D_\mu -  m_B) B  \nonumber \\
   & &   + \; F\, \Tr \bar{B}\gamma^\mu \gamma_5 [ {\mathcal  A}_\mu,B]
  + \; D\, \Tr \bar{B}\gamma^\mu \gamma_5 \{ {\mathcal  A}_\mu,B\}
\label{lan-KN}
\\
   & & + c \Tr {\mathcal   M} (U + U^\dagger) +
   a_1 \Tr \bar{B}(\xi {\mathcal  M} \xi + \xi^\dagger {\mathcal  M}
\xi^\dagger) B
\nonumber \\
   & & + a_2 \Tr \bar{B}B(\xi {\mathcal  M} \xi + \xi^\dagger {\mathcal
 M} \xi^\dagger)
+
   a_3 \Tr \bar{B}B\; \Tr(\xi {\mathcal  M} \xi + \xi^\dagger {\mathcal
M} \xi^\dagger)
.\nonumber
\eeqa
The Lagrangian consists of one part which is $SU(3)\times SU(3)$
symmetric (the first four terms in (\ref{lan-KN})), and the other part
which breaks the chiral symmetry, where the breaking strength is
determined by the parameters
$a_1, a_2, a_3$, and $c$, as well as by the quark mass matrix
$\mathcal M$ which is
\beq
\mathcal M =  \left(
  \begin{array}{ccc}
  0 & 0 & 0 \\
  0 & 0 & 0 \\
  0 & 0 & m_s
  \end{array}
  \right){,}
\eeq
i.e. the light quarks are assumed to be massless. Parameters in the
symmetric part are $f = 93 $ MeV, $D=0.81, F=0.44$. In the
symmetry-breaking sector $c$ represents
the vacuum expectation value of the scalar quark density $\bar{q}q$
\cite{Weinberg:1996kr}. The Gell-Mann-Oakes-Renner
relation states that $m_K^2  = 2 c m_s /f^2$.
Mass splittings for the strange baryons determine the values of $a_1
m_s, a_2 m_s$  and we use after \cite{Politzer:1991ev}
$a_1 m_s = -67 $ MeV and $a_2 m_s = 134 $ MeV.
The last parameter $a_3 m_s$ is subject to a large uncertainty. In
principle it should be extracted from
the kaon-nucleon scattering or kaonic atoms data \cite{Lee:1996ef}
through the  kaon-nucleon sigma term
$\Sigma_{KN} = \frac{1}{2} m_s \langle N | \bar{u} u + \bar{s}s | N
\rangle =
 -(\fract{1}{2}a_1 + a_2 + 2 a_3) m_s $ ,
but up to now its value has  not been determined. To assess its
influence on further results we adopt its values in the range
$-310 \ \mathrm{MeV} < a_3 m_s < -134  \ \mathrm{MeV}$
which corresponds to a reasonable strangeness content of the proton
in the range:
\mbox{$0.2 > {\langle \bar{s}s \rangle_p} / {
\langle \bar{u}u + \bar{d}d + \bar{s}s\rangle_p} > 0
$} \cite{Donoghue:1985bu}.
The value of $a_3 m_s$ appears to be important for the kaon condensation
because the kaon-nucleon sigma term provides the attractive interaction
between kaons and nucleons, and makes the kaon effective mass smaller in
the nuclear medium. Negatively charged kaons condense in the
 nucleon matter if their effective mass becomes smaller then the
electron chemical potential $\mu_e$. Assuming that the only relevant
kaon-nucleon interaction is in the s-wave,
the  expectation value of the kaon field $\langle K^- \rangle$ is
spatially uniform. According to the Baym theorem \cite{baym-theorem},
its time dependence is
\beq
\langle K^- \rangle = \frac{f \th}{\sqrt{2}} \exp(-\mu_K t),
\eeq
where $\th$ is an unknown amplitude of the condensate.

\begin{figure}
\center{\includegraphics[height=2.5in]{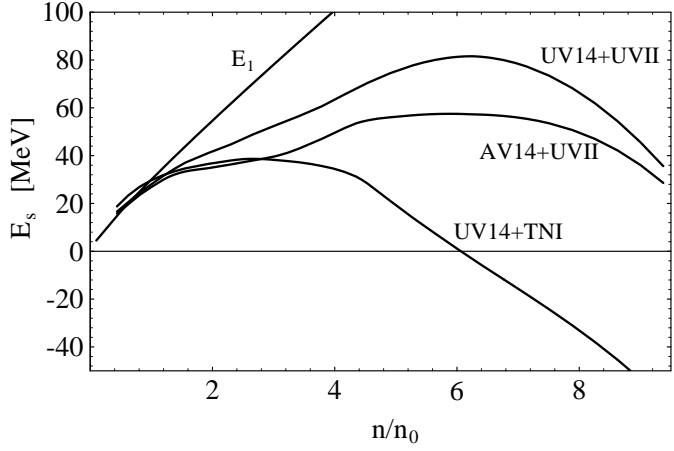}}
\caption{\small The nuclear symmetry energy  for different realistic
potential models: UV14+UVII, AV14+UVII, and UV14+TNI. The line labelled
$\mathrm{E_1}$ corresponds to the symmetry energy for the RMF model
\cite{Thorsson:1994bu}.}
\label{rys_Es}
\end{figure}

Formally, pure nucleon-nucleon interactions should also emerge
from the chiral theory as it is supposed to
represent the low energy limit of QCD. However, it was not derived up to
now, so
we use the chiral Lagrangian only to extract the kaon-nucleon part of
the interactions and we adopt the nucleon-nucleon part from the
realistic potential models of nuclear matter.
This means that the energy density of our problem is a sum of three
contributions: the kaon-nucleon interaction energy density (derived
from ${\mathcal L}_\chi$ restricted only to kaon  and
nucleon fields), the nucleon-nucleon energy density taken from
\cite{Wiringa:1988tp} and the energy density for leptons ($e, \mu$):
\beq
\ep = \ep_{KN} + \ep_{NN} + \ep_{lep}.
\label{en-tot}
\eeq
The nucleon-nucleon contribution is
\beq
\ep_{NN} = \frac{3}{5} E_F^{(0)} n_0 \left(\frac{n}{n_0}\right)^{5/3}
          + V(n) + m n +  n(1 - 2x)^2 E_s(n),
\label{endens-NN}
\eeq
where
\beqa
V(n)&= & n V_0(n) ,\\
E_s(n)& =& \frac{1}{3} E_F^{(0)} \left(\frac{n}{n_0}\right)^{2/3} + V_2(n) .
\eeqa
The functions $V_0$ and $V_2$ depend on the parameterization
of the nucleon-nucleon potential used to fit the experimental data. We
made use of the functions  given in \cite{Wiringa:1988tp} corresponding
to UV14+UVII, AV14+UVII, UV14+TNI interactions, where
UV14 and AV14 are 14-operators two-body potentials from Urbana and
Argonne. The three-body potentials UVIII or TNI are important to
reproduce the saturation point properties by providing a proper
contribution in the isoscalar sector. The UV14+TNI potentials reproduce
best the saturation density $n_0 = 0.16~\mathrm{fm^3}$ and the binding
energy $w_0 = 16~\mathrm{MeV}$. One can notice that the three-body forces
give different energies also in the isovector sector, i.e. the
symmetry energy, as shown in Fig.\ref{rys_Es}.
For the same two-body potential, UV14, the three-body potentials UVII
and TNI give different energies at higher densities,
although both of them produce the same
symmetry energy at saturation point $E_s(n_0) \approx 30$ MeV.
{In this figure, also the symmetry energy, typical for RMF theory, is shown.
It is given by the formula \cite{Thorsson:1994bu}
\beq
E_s(n) = \frac{3}{5}(2^{2/3} - 1)  E_F^{(0)} (u^{2/3}  - F(u)) 
  + E_s(n_0) F(u) ,
\eeq
where $u=n/n_0$ and the interaction contribution $F(u)$ depends linearly on 
density, $F(u) = u$.}
The nuclear symmetry energies from realistic potential models differs
from one another but only quantitatively,
whereas the RMF prediction $E_1$ is qualitatively different - it never
saturates at higher densities.

The energy density for leptons  is determined in a standard way through
their Fermi momenta
\beq
\ep_{lep} = \frac{k_e^4}{4 \pi^2} + m_\mu^4
g(\sqrt{\eta(k_\mu^2-m_\mu^2)}/m_\mu)~,
\eeq
where the function $g(t)$ comes from integration over the Fermi sea of
massive  fermions:
\beqa
g(t) & = & \frac{1}{8 \pi^2} ((2 t^3 + t) \sqrt{1 + t^2} -
\mathrm{arsinh} t )~,
\\
& & \eta(x) = \left\{ \arl x ~;~ x \geq 0 \\  0 ~;~ x \leq 0~~~. \earl
\right.
\nonumber
\eeqa
To derive the kaon-nucleon contribution, $\ep_{KN}$, in (\ref{en-tot}) we
use the standard prescription  to get the energy density from a given
Lagrangian (Appendix \ref{app-a1}). Employing the Baym theorem we
obtain finally
\beqa
\ep_{KN} &=& f^2 {\mu_K^2\over2} \sin^2 \th +
     2 m^2_K f^2 \sin^2 {\th\over2} + n (2a_1 x + 2a_2 + 4a_3) m_s
\sin^2{\th\over2}~.
\label{endens-KN}
\eeqa

\section{Thermodynamics of the system}
\label{sec_thermod}

\begin{figure}
\center{\includegraphics[height=4.0in]{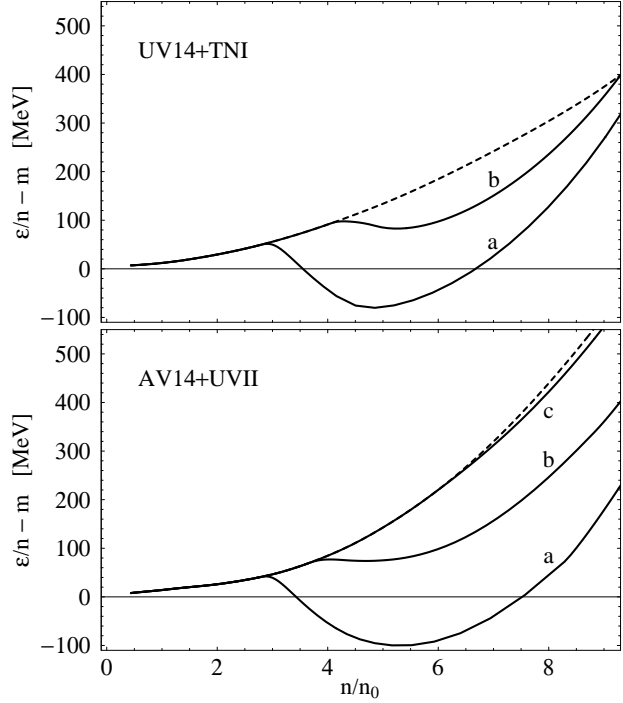}}
\caption{\small Energy per particle for kaon-condensed neutron star
matter
with UV14+TNI (upper panel) and AV14+UVII (lower panel) interactions.
The
labels a, b and c correspond to the values of $a_3m_s$ from 
Table \ref{tab-nc}.}
\label{rys_Ene}
\end{figure}

At zero temperature  the energy density should  be  canonically a
function of the number densities $\ep(n_p,n_n,n_K)$ (leptons may be
disregarded without loss
of generality). Instead of  number densities we have  $n, x, \mu_K$ in
the expressions (\ref{endens-KN}),(\ref{endens-NN}) entering $\ep$, which
describe the thermodynamic state completely. The fourth,
superfluous variable, $\th$, is
a free microscopic parameter with respect to which the energy is
minimized when all densities are fixed,
\beq
\left(\frac{\pa \ep}{\pa \th}\right)_{n_i} = 0.
\eeq
This condition leads to the equation for $\th$ (Appendix \ref{app-a2})
which now may be expressed in terms
of the equilibrium state determined by $n,x,\mu_K$,
\beq
\cos\th = \frac{1}{f^2 \mu_K^2} \left(m_K^2 f^2 + \fract{1}{2}n(2a_1
x\!+\!2a_2
\!+\!4a_3)m_s
    - \fract{1}{2}\mu_K n (1\!+\!x) \right)~.
\label{th-min}
\eeq
The function $\th(n,x,\mu_K)$ plays the role of the order parameter in
the analysis of the phase transition.
The kaon condensate develops provided the above equation possesses a
solution. If there is no solution, i.e.
the right-hand side of eq.(\ref{th-min}) does not fall in the range
(-1,1), this means dense matter is allowed to contain only the normal
nucleon phase. In the neutron star matter the free thermodynamic
parameters $n,x,\mu_K,\mu_e,\mu_\mu$ are subject to further
constraints which come from the beta equilibrium and charge neutrality.

On the timescale relevant to cooling of neutron stars, the nucleon
matter with a kaon condensate approaches the beta equilibrium through the
following reactions :
\beqa
n &\leftrightarrow & p + l + \nu_l \nonumber\\
n &\leftrightarrow & p + K^- ~,\label{beta} \\
l & \leftrightarrow & K^- + \nu_l~~,~~~l = e, \mu \nonumber
\eeqa
As neutrinos are supposed to leave freely, from the above reactions one
derives corresponding
relations among chemical potentials for all
particles present in the system:
\beq
\mu_K = \mu_e = \mu_\mu =  \mu_n - \mu_p
\label{beta-seq}
\eeq
The above sequence of equalities shows that there are two independent
chemical potentials, let's say $\mu_e$ and $\mu_n$. These chemical
potentials represent two quantities, the electric charge and the
baryon number, which are conserved in the beta-reactions (\ref{beta}).
 This means that the charge and the baryon number
uniquely describe the thermodynamic state of our system. In further
calculations we put
\[
\mu \equiv \mu_e~,
\]
and $\mu$ will represent the  chemical potential of all negatively
charged particles.
Chemical potentials for leptons are determined by their Fermi momenta,
$\mu_e = k_e$, and, $\mu_\mu = \sqrt{k^2_\mu + m_\mu^2}$,
and for nucleons they must be derived through a proper differentiation,
$\mu_{p,n} = (\pa \ep / \pa n_{p,n})_{n_i} $, i.e. with all remaining
densities  fixed:
\beqa
\mu_n\!&\!=\!&\! E_F^{(0)}\left(\frac{n}{n_0}\right)^{2/3} + m + V'(n) +
n(1\!-\!2x)^2 E_s'(n) + (1\!-\!4x^2)E_s(n)  \nonumber \\
&&  +\;(2a_2m_s + 4 a_3 m_s - \mu_K) \sin^2{\th\over 2} ,
\label{mun} \\
\mu_p\!&\!=\!&\! E_F^{(0)}\left(\frac{n}{n_0}\right)^{2/3} + m + V'(n) +
n(1\!-\!2x)^2 E_s'(n) - (1\!-\!2x)(3\!-\!2x)E_s(n) \nonumber \\
&&  +\;(2 a_1 m_s  + 2a_2m_s + 4 a_3 m_s - 2\mu_K) \sin^2{\th\over 2} .
\label{mup}
\eeqa
With these expressions for $\mu_p, \mu_p$,  the beta equilibrium
condition
(\ref{beta-seq})
leads to the equation
\beq
\mu  =  4(1-2x) E_s(n)/ \cos^2\frac{\th}{2} - 2 a_1 m_s
\tan^2\frac{\th}{2}.
\label{k-beta}
\eeq
While the beta equilibrium condition is easily implemented, the
assumption as to the charge neutrality needs further qualification.
 We must make distinction between local
and global electric neutrality of the system. For locally neutral
matter in any point densities of charged particles satisfy the
 condition
\[
n_p - n_K - n_e  - n_\mu = 0.
\]
The only unknown quantity in the above formula is the negative kaon
density
$n_K$,
which is obtained from the conserved electric current corresponding to
the
$U(1)$ symmetry
of the kaon-nucleon Lagrangian
\[
n_K = i (p_{K^+} K^+ -  p_{K^-} K^- ),
\]
where $p_{K^+}, p_{K^-}$ are the canonical momenta.
After somewhat laborious calculations they may be derived directly
from the Lagrangian (Appendix \ref{app-a1}). Making use of the Baym
theorem we get the final expression for the charge density carried by
kaons,
\beq
n_K =  f^2 \mu \sin^2\th +  n (1+ x) \sin^2{\th\over2} ,
\label{kaon-dens}
\eeq
and the local charge neutrality condition becomes:
\beq
0  =  f^2 \mu \sin^2\th +  n (1+ x) \sin^2{\th\over2}
    +\frac{\mu^3}{3 \pi^2} +
       \th(\mu)\frac{(\mu^2 - m_\mu^2)^{3/2}}{3\pi^2}  - nx~.
\label{neutr-local}
\eeq

When a phase transition takes place, the charge separation
in two different phases preserving merely the neutrality of the matter
as a whole, but not on a small scale, can occur, as it was shown
in \cite{Glendenning:1992vb}. Then the equation (\ref{neutr-local})
does not longer hold.
The problem of mixed phase formation will be discussed elsewhere.
\begin{figure}
\center{\includegraphics[height=2.5in]{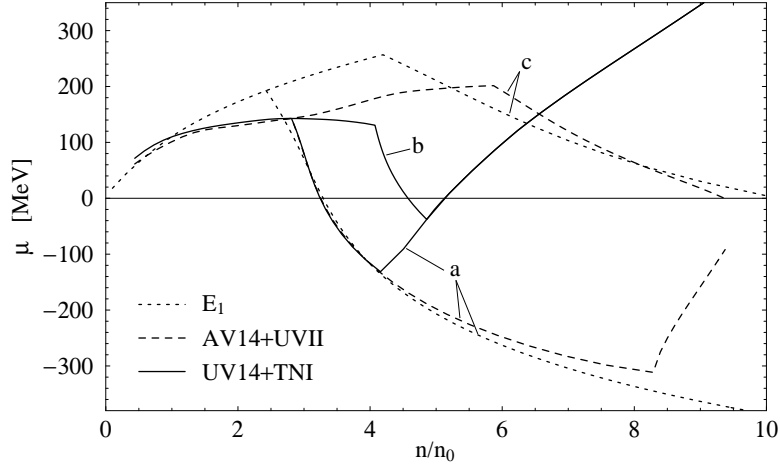}}
\caption{\small The kaon chemical potential for indicated
nuclear models and different values of $a_3m_s$ (models a,b and c from
 Table \ref{tab-nc}).
\label{rys_mu}}
\end{figure}

The equations  (\ref{neutr-local},\ref{k-beta},\ref{th-min}) allow us
to find the total energy density (\ref{en-tot}) as a function of
 the baryon density  only, $\ep(n)$.
The onset of condensation occurs at a critical density $n_c$  where the
function $\th$ starts
to deviate from zero and the energy density of the phase with the
condensate becomes smaller than that
of normal matter, Fig.\ref{rys_Ene}. As it was shown in \cite{Kubis:1999sc},
 the critical density, $n_c$, depends sensitively on
the shape of the symmetry energy, $E_s(n)$, with the eventual exclusion of
condensation if the strangeness is too small.
The values of $n_c$ for different models of the symmetry energy, $E_s$,
and some representative values of strangeness  are presented in
 Table \ref{tab-nc}.
\begin{table}[ht]
\caption{\small Critical density for the kaon condensation in units of $n_0$.
In case of UV14+TNI interactions with the minimal strangeness
$a_3 m_s = -134 \mathrm{MeV}$ the kaon condensate does not appear.}
\label{tab-nc}
\begin{center}
\begin{tabular}{|c|c|c|c|c|c|}
\hline
  &  $a_3 m_s~[\mathrm{MeV}]$ & $~{\mathrm{E_1}}~$ & UV14+UVII & AV14+UVII &
UV14+TNI \\
\hline
a & -310 & 2.4 & 2.6 & 2.7 & 2.8 \\
b & -230 & 3.0 & 3.4 & 3.7 & 4.1 \\
c &  -134 & 4.2 & 5.0 & 5.9 &  -- \\
\hline
\end{tabular}
\end{center}
\end{table}
Likewise,  the other quantities, important for the structure of
neutron stars, change radically with $E_s$. Paradoxically,
low values of the symmetry energy (which usually reduce $x$)
in the presence of the condensation lift the proton abundance
even up to a complete protonization, $x=1$. In the most extremal case
of the UV14+TNI interactions the protonization sets in just after
 formation of the condensate.
This is because kaons appear at  a small value of
the kaon chemical potential $\mu$, that means they may be easily
produced as the chemical potential measures the cost of energy to add
one particle to the system. The kaon chemical potential in the
condensate is shown in Fig.\ref{rys_mu}
for some models we study here.

\begin{figure}
\center{\includegraphics[height=2.5in]{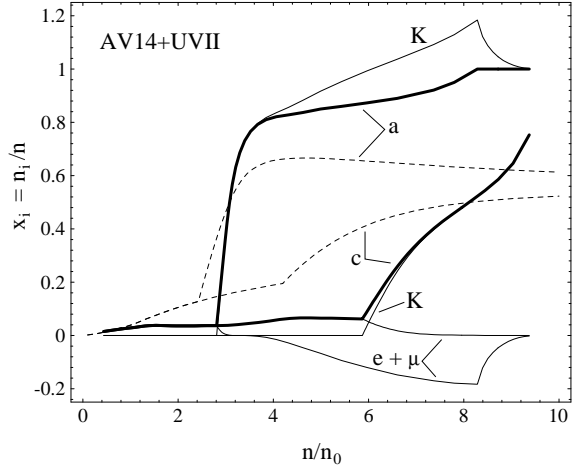}}
\caption{\small The composition of the kaon-condensed neutron star matter  for
AV14+UVII potentials.
Two families of curves indicated as (a) and (c) correspond to maximal
and zero strangeness  in Table \ref{tab-nc}, respectively. Bold lines
represent the proton fraction $x$, thin lines $x_K$ and $x_{e+\mu}$.
For comparison, dashed lines correspond the proton fraction $x$ calculated
with the same function $V_0$ and the $E_1$ model of the nuclear symmetry
symmetry energy.}
\label{xxallav.eps}
\end{figure}

\begin{figure}
\center{\includegraphics[height=2.5in]{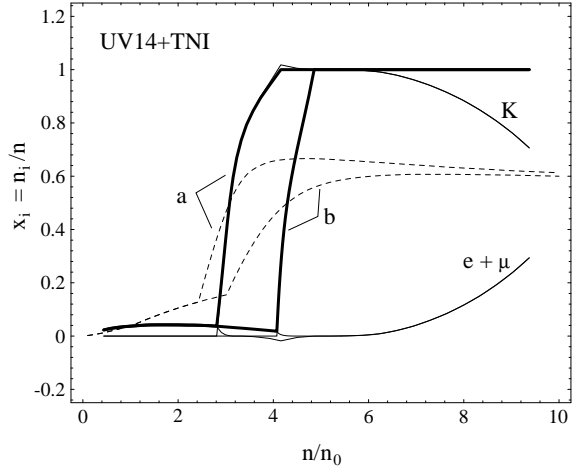}}
\caption{The composition of the kaon-condensed neutron star matter for
UV14+TNI potentials.
Line labels as in  Fig.\ref{xxallav.eps}.
For zero strangeness (c) there is
no condensation, so only (a) and (b) families are presented. At high
density all particle fractions $x_i$ are independent of the strangeness
and corresponding curves overlap for  different families.}
\label{xxalltni.eps}
\end{figure}

With the presence of kaon condensate the overall chemical composition of
the neutron star matter becomes somewhat exotic. This is shown in
Fig.\ref{xxallav.eps}
 and \ref{xxalltni.eps}. Let us consider first the case of
AV14+UVII (UV14+UVII gives similar results but
less pronounced). Kaons, as bosons, are produced in the zero momentum
state unlike leptons which must be put on the top of the Fermi sea.
 The negative charge carried
by kaons is so abundant that the kaons supersede electrons and muons in
dense matter - the curves for $x$ and $x_K$ overlap
(case (c) in Fig.\ref{xxallav.eps}).  When
strangeness is high (case (a) in Fig.\ref{xxallav.eps})
 the kaons are produced at lower
density, $x_K$ approaches the value $x_K=1$ and even may exceed unity.
If $x_K$ is so high, positive leptons ($x_{e+\mu} < 0$) have to appear
 to ensure  charge neutrality.
With further increase of density the situation changes again.
As the proton fraction $x$ cannot be greater than 1, the equation
(\ref{th-min})
is no longer valid. Strictly speaking, the energy has no minimum for $x$
inside the allowed range, $x \in (0,1)$. So, it
takes its minimal value at the edge of the range of $x$ which is just at
$x=1$. Then $\mu$ stops to decrease, positive leptons disappear and
negative leptons may be restored. This fact is better  seen  in
Fig.\ref{xxalltni.eps}.
In the UV14+TNI case  positive leptons appear as well but with much
 smaller amount because
the effect of $x=1$ constraint is very strong. And then, at higher
density, the composition
of matter does not depend on the strangeness value. The curves for
$x_K$ and $x_{e+\mu}$ overlap for high $n$ (see Fig.\ref{xxalltni.eps}).

The above discussion clearly indicates  the situation becomes complex.
There is a common conviction that high proton fraction $x \gtrsim 0.15$
 ensures very fast cooling of neutron stars by direct URCA processes
\cite{Lattimer:1991ib}.
This statement is true only for usual $n,p,e,\mu$ matter. Direct URCA
cycles take place if Fermi momenta for protons, neutrons and leptons
form a closed triangle.
All the Fermi momenta may be expressed in terms of $x$. So, the
triangle condition leads to the existence of the threshold value,
$x_\mathrm{URCA}$, which lies between 0.11 and 0.15, depending on the
presence of muons. For nuclear matter with kaons we have an
additional degree of freedom -  $\th$,  which determines the kaon
abundance altogether with $\mu$ (\ref{kaon-dens}).
This is the reason why we cannot formulate any simple answer to the
question if direct URCA processes are allowed above some value of
 $x$ or not.
In our case, the triangle condition must be examined explicitly in
 the whole range of densities.

\section{Cooling of matter with kaon condensate}
\label{sec_cooling}

\begin{figure}
\center{\includegraphics[height=4in]{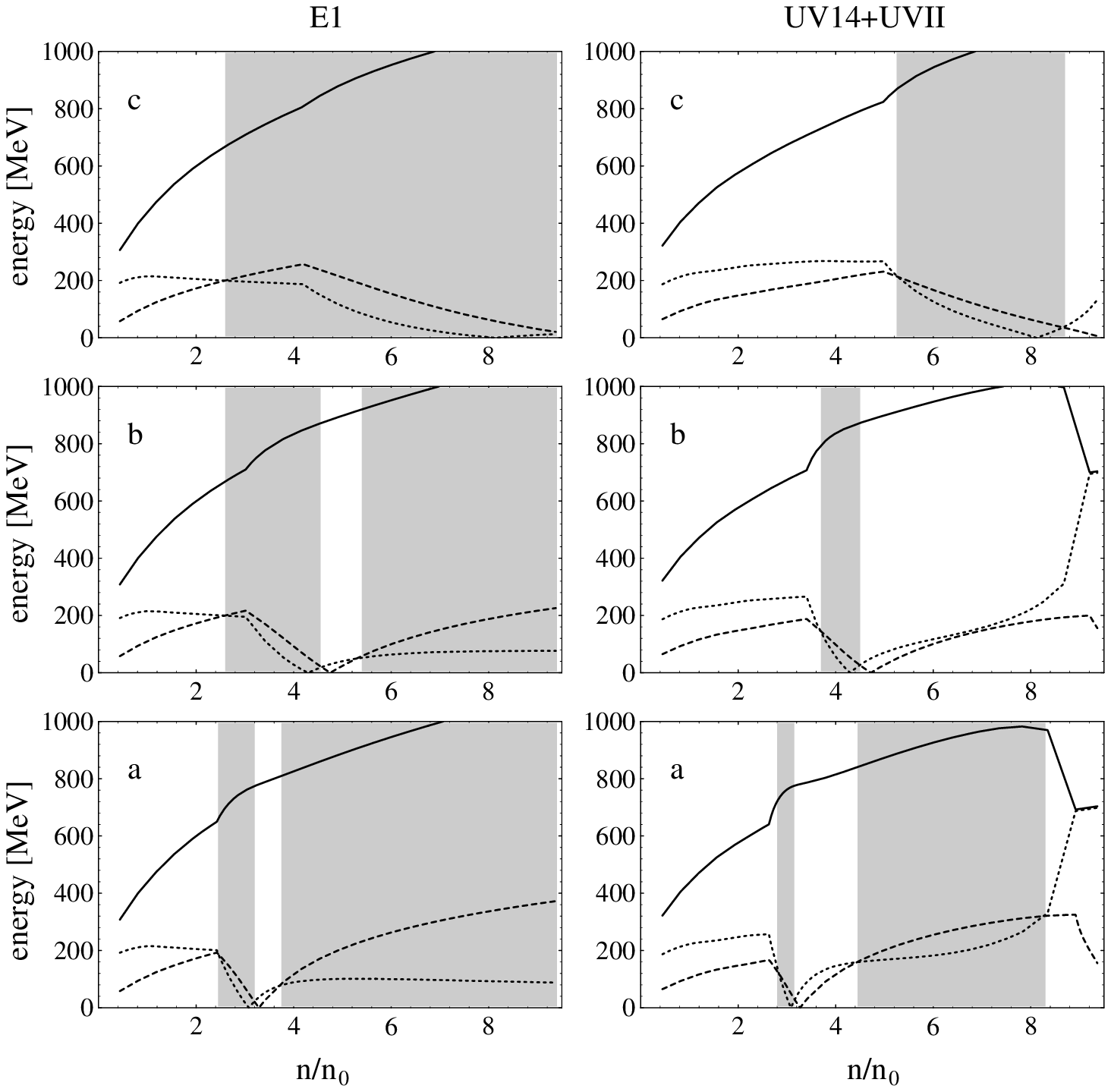}}
\center{\includegraphics[height=4in]{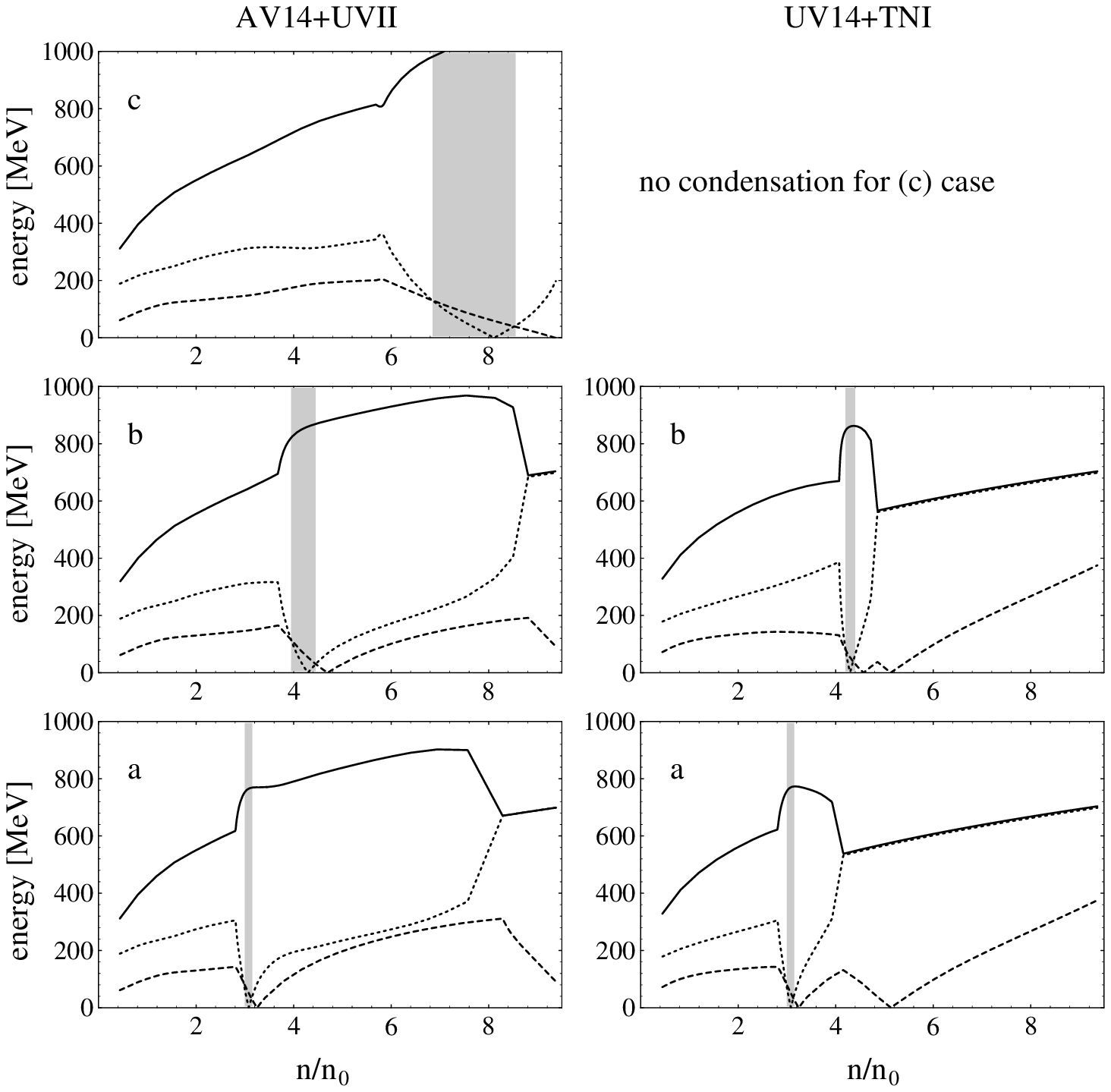}}
\caption{The dURCA triangle condition for various nuclear models and
different values of strangeness. The solid lines correspond to
$k_{\tilde{p}} + k_{\tilde{n}}$, dashed lines to $k_e$ and dotted lines
to
$|k_{\tilde{p}} - k_{\tilde{n}}|$. The shaded areas represent regions
where
dURCA is allowed}
\label{urcall}
\end{figure}

\begin{figure}[t]
\center{\includegraphics[width=4.5in]{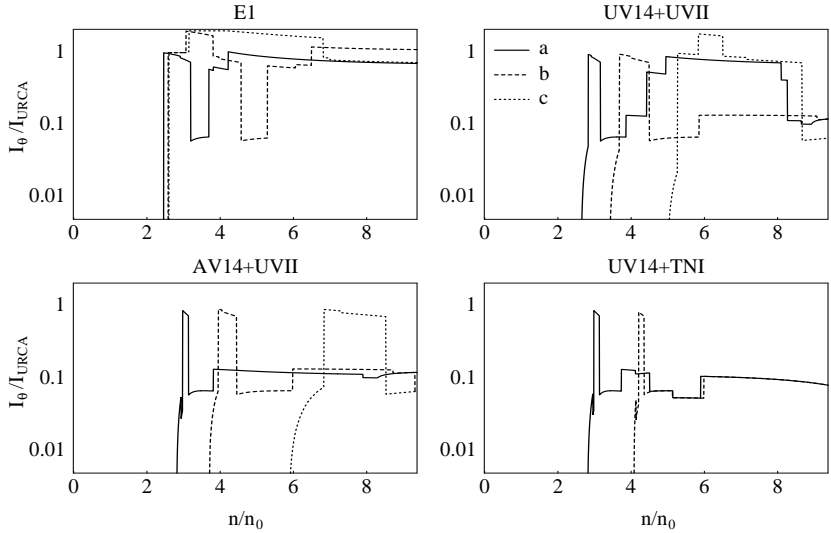}}
\caption{The total neutrino emissivity in URCA processes for various
nuclear models.}
\label{intallmu.eps}
\end{figure}

When a kaon condensate is formed, besides the usual direct beta
reactions there are also possible reactions which are specific to the
kaon-condensed neutron star matter.
The direct URCA cycles take place among quasi-particles
$\tilde{p}, \tilde{n}$
\beqa
 \tilde{n} \rightarrow  & \tilde{p} + l &  +~\bar{\nu}_l  \nonumber \\
 & \downarrow &   \label{k-dURCA} \\
 & \tilde{p} + l & \rightarrow \tilde{n}  + \nu_l ,
\nonumber
\eeqa
and we call them dURCA.
The states $|\tilde{p} \rangle , |\tilde{n}\rangle$ are linear
combinations of pure proton and neutron state which depend on the
condensate  amplitude $\th$ \cite{tatsumi-cool}. As the
 quasi-particles have no definite charge
one may write another admitted type of reactions
\beqa
\tilde{n} & \rightarrow  & \tilde{n} + l   +~\bar{\nu}_l
\label{kURCA-n}, \\
\tilde{p} & \rightarrow  & \tilde{p} + l   +~\bar{\nu}_l~.
\label{kURCA-p}
\eeqa
Microscopically, it means that the necessary charge is taken from the
condensate, e.g. $n + K \rightarrow n + l + \nu_l$  and thus they are
called {\em kaon induced} URCA.  For all particles taking part in the
above reactions, the momentum conservation gives the following relations
among their Fermi momenta:
\beqa
\mathrm{kURCA: }& &~~~~~~
2 k_{\tilde{n}}>k_l~~~\mathrm{or}~~~2 k_{\tilde{p}}>k_l \\
\label{triangle-k}
\mathrm{dURCA: }& &~~~~~~ |k_{\tilde{p}} - k_{\tilde{n}}| < k_l <
k_{\tilde{p}} +
k_{\tilde{n}}~.
\label{triangle-d}
\eeqa
The neutrino momentum, which is $k_\nu \sim kT < 1 \mathrm{MeV}$, may be
neglected as it is much smaller than momenta of the other particles.
Likewise, kaons do not enter into the relation (\ref{triangle-k}) as the
condensation takes place in the s-wave and the kaon momenta do not
contribute.
Let's focus on the direct URCA cycle. The three terms of the double
inequality  are presented for various models in Fig.\ref{urcall}.
When negative leptons are present, the right-hand side inequality
in (\ref{triangle-d}) is always fulfilled , because the local
neutrality means that $k_{e} \le k_{\tilde{p}} $.
It does not have to be true when positive
leptons appear. In principle, although very unlikely, it is possible
that $k_{e} > k_{\tilde{p}}$  and for clarity we
put the sum $k_{\tilde{p}} + k_{\tilde{n}}$ on the plots as well.
The main conclusion from the calculations shown in Fig.\ref{urcall} is
that the exceeding  by the  proton fraction
the threshold  value $x_\mathrm{URCA}$ does not ensure the presence of the
dURCA cycle. Too high values of  $x$ block this cycle due to small
fraction of neutrons.
Also, a non-typical behaviour of the chemical potential $\mu$, which
crosses zero at some density (Fig.\ref{rys_mu}), suppresses dURCA
 in a narrow zone due to small lepton abundance. From the plots in
Fig.\ref{urcall} the tendency of suppressing the dURCA process with
smaller values of the symmetry energy can be inferred. However, the
importance of this effect is not so great, because always one of
the kaon-induced processes (with a proton or a neutron) operates
as may be easily concluded from Fig.\ref{urcall}.
The sum $k_{\tilde{p}}  + k_{\tilde{n}}$ is always greater than $k_e$
so, either $2k_{\tilde{p}} > k_e$ or $2 k_{\tilde{n}} > k_e$ or both
 of them are fulfilled.
\begin{table}[th]
\begin{center}
\caption{Emissivities of the three fermion processes in the kaon-condensed
neutron star matter.}
\label{tab-urca}
\vspace{1em}
\begin{tabular}{|c|c|c|}
\hline
cycle   & reaction & $I_\th/I_\mathrm{URCA}$ \\
\hline
dURCA & $\tilde{n} \rightarrow \tilde{p} + l + \bar{\nu}_l $ &
$\cos^2{(\th/2)}$
\\
\hline
kURCA & $\tilde{n} \rightarrow \tilde{n} + l + \bar{\nu}_l $ &
$\frac{1}{4}\sin^
2{\!\th}\;\tan^2{\th_c} $\\
      & $\tilde{p} \rightarrow \tilde{p} + l + \bar{\nu}_l $ & $
\sin^2{\!\th}\;
\tan^2{\th_c} $\\
\hline
\end{tabular}
\end{center}
\end{table}
In the thermal evolution of neutron stars  we distinguish two
cooling  scenarios: {\em fast} -- driven by direct URCA  and
{\em slow} -- driven by modified URCA processes. Modified URCA
always operate, as they are not limited by any kinematical constraints,
however their emissivities are much smaller, about 5 to 6 orders
of magnitude less than the direct ones.
So they provide much slower cooling.
The kaon-induced URCA processes (like direct URCA) belong to the class of
three-fermion processes but their emissivity
is smaller because they proceed through strangeness-changing reactions.
The rate of such reactions scales as $\sin\th_c$, whereas direct URCA
scales as $\cos\th_c$, where $\th_c$ is the Cabibbo angle. Both kinds of
 processes depend on the
kaon abundance which is described by the condensate amplitude $\th$
\cite{Thorsson:1995rk}.
In Table \ref{tab-urca}
the ratio  of emissivities of kaon-condensed neutron star matter to
emissivity by direct URCA of normal matter with no condensate
is presented.
The emissivity finally depends on the baryon number density through
$I_\mathrm{URCA}$
and the condensate amplitude $\th$. The total emissivity is a sum of
all the contributions from direct and kaon-induced URCA cycles with
electrons as well as muons which are present if $\mu_e > m_\mu$.
In Fig.\ref{intallmu.eps}  the total emissivity is shown.

\section{Summary and discussion}

We have shown that the formation and properties of the kaon-condensed
phase of neutron star matter are quite sensitive to the high density
behaviour of the nuclear symmetry energy, $E_s(n)$, which still remains
the most poorly known property of dense matter. In particular, for
$E_s$ decreasing to negative values the formation of the
kaon condensate can be inhibited for the lowest
absolute value of $a_3m_s$. This is in stark contrast to the
 results  for monotonically increasing $E_s(n)$, when the condensation
occurs for any value of $a_3m_s$.
The properties of the kaon-condensed neutron star matter and the
abundance of particle species are sensitive to the high density form of
the nuclear symmetry energy, as shown in
Figs.\ref{xxallav.eps}-\ref{xxalltni.eps}. A surprising
finding is that the direct URCA process is often not allowed in the
 kaon-condensed matter for decreasing $E_s(n)$.

The above results allow us to assess how great is the influence of
nuclear force models on the total neutrino emissivity of kaon-condensed
neutron star matter. In case of increasing symmetry energy as in the
RMF model $\mathrm{(E_1)}$ dURCA operates almost always when kaon condensate
is formed, except of a narrow zone around density where
the chemical potential $\mu$ is zero. For realistic potential models
dURCA is systematically suppressed  to more narrow density range with
decreasing symmetry energy. For an extreme case of the UV14+TNI
interactions, the allowed zone shrinks
eventually to a very narrow vicinity of some density close to the
threshold value for condensation $n_c$. The kaon-induced URCA branch
 is always present but
at the level of about 10 times smaller, as shown in
Fig.\ref{intallmu.eps}. 

\begin{figure}[h]
\center{\includegraphics[height=3in]{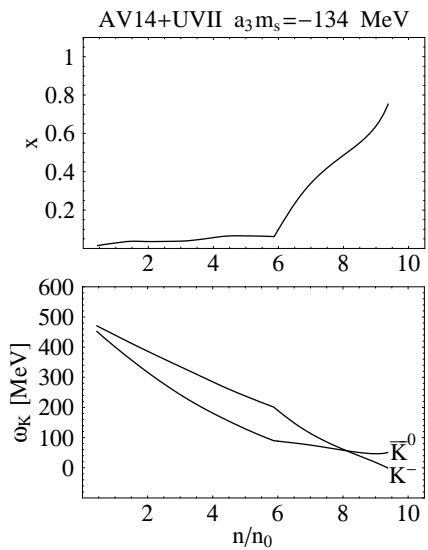}}
\caption{The proton fraction $x$ and anti-kaons energy in dense matter.
The leveling off of $\omega_K$ for ${\bar{K}^0}$ corresponds to the onset of 
$K^-$ condensation which drives the strong increase of the proton fraction 
shown in the upper panel.}
\label{omega_rys}
\end{figure}

Another interesting  question concerns the role of $\bar{K}^0$.
As was shown in \cite{Pal:2000pb} neutral kaons are easily produced
just after the onset of $K^-$ condensation. Then the kaon-condensed matter 
becomes
even more effectively isospin-symmetrized than in the absence of $\bar{K}^0$. 
We would like to stress that this effect is sensitive to the form of the 
nuclear symmetry energy. The effect occurring for increasing $E_s(n)$ is not 
observed in the case of 
decreasing $E_s(n)$. High values of $x$ which we obtain in our model
make the $\bar{K^0}$ condensation  more difficult. The dispersion relations 
for anti-kaons are:
\beq
\omega_{K^-} = - \frac{n(1+x)}{4 f^2} + 
 \left(\left(\frac{n(1+x)}{4 f^2}\right)^2 + m_K^2 +  
    (2a_1 x + 2a_2 + 4 a_3) m_s \frac{n}{2f^2}  \right)^{1/2} 
\eeq
\beq
\omega_{\bar{K}^0} = - \frac{n(2-x)}{4 f^2} + 
 \left(\left(\frac{n(2-x)}{4 f^2}\right)^2 + m_K^2 +  
    (2a_1 (1-x) + 2a_2 + 4 a_3) m_s \frac{n}{2f^2}  \right)^{1/2}
\label{omega_k0}
\eeq
The first term in the expression (\ref{omega_k0}) for $\omega_{\bar{K}^0}$,
which is the leading term, includes 
positive contribution from $x$, so high values of $x$  make the slope of
$\omega_{\bar{K}^0}$, as a function of $n$, more flat after the charged kaons 
appear.
Fig.\ref{omega_rys} presents this effect clearly. We leave more detailed 
discussion of the neutral kaon condensate for future work. 

Note added: After completion of this work dr. T. Muto brought to our attention 
his research \cite{Fujii:ua} concernig  kaon condensation and cooling of 
neutron stars.
\newpage

\appendix

\section{Calculation of thermodynamic quantities}

\subsection{Basic formulae}
\label{app-a1}
A matrix $U \in SU(3)$ is defined as $U =\exp(i\sqrt{2}\fracd{M}{f})$,
where $M$ represents the pseudoscalar meson octet
\beq
M = \left(
  \begin{array}{ccc}
  \frac{1}{\sqrt{2}} \pi^0 + \frac{1}{\sqrt{6}} \eta_8 & \pi^+ & K^+ \\
  \pi^- & -\frac{1}{\sqrt{2}} \pi^0 + \frac{1}{\sqrt{6}} \eta_8 & K^0 \\
  K^- & \bar{K}^0 & -\sqrt{\frac{2}{3}} \eta_8
  \end{array}
    \right).
\eeq
Matrix $\xi$ appearing in (\ref{lan-KN}) is such that $\xi^2 = U$ and
$B$ corresponds to the baryon octet
\beq
B = \left(
  \begin{array}{ccc}
  \frac{1}{\sqrt{2}} \Sigma^0 +  \frac{1}{\sqrt{6}} \Lambda & \Sigma^+
& p \\
  \Sigma^- & -\frac{1}{\sqrt{2}} \Sigma^0 + \frac{1}{\sqrt{6}} \Lambda
& n \\
  \Xi^- & \Xi^0 & -\sqrt{\frac{2}{3}} \Lambda
  \end{array}
    \right).
\label{baryons}
\eeq
When the only non-zero elements of the meson  matrix $M$ are $K^+$ and
$K^-$ fields, then $U$ has the form
(below we use dimensionless meson fields $k^\pm = \sqrt{2} K^\pm/f$):
\beq
U = \left(
\begin{array}{ccc}
\cos{\sqrt{k^- k^+}} & 0 &
\frac{i  k^+ \sin ({\sqrt{k^- k^+}})}{{\sqrt{k^- k^+}}} \\
 0 & 1 & 0 \\
 \frac{i  k^- \sin ({\sqrt{k^- k^+}})}{{\sqrt{k^- k^+}}} & 0 &
 \cos ({\sqrt{k^- k^+}})
\end{array}
\right).
\eeq
Hence the canonical momenta corresponding to the rescaled kaon fields
$p_\pm = \fracd{\pa {\mathcal L}}{\pa\, (\pa_t k^\pm)}$ are
\beqa
p_+ & = & -i\, (\bar{n}n + 2 \bar{p}p) \frac{\sin^2\frac{\chi}{2}}{2
\chi^2}
k^-
 \nonumber \\
 & & + \; \frac{f^2}{4}
   \left(\frac{k^+ \pa_t k^- + k^- \pa_t k^+}{k^+} +
     \frac{k^+ \pa_t k^- - k^- \pa_t k^+}{k^+}
\,\frac{\sin^2\chi}{\chi^2}
   \right) \\
p_- & = & ~i\, (\bar{n}n + 2 \bar{p}p) \frac{\sin^2\frac{\chi}{2}}{2
\chi^2}
k^+
 \nonumber \\
 & & + \; \frac{f^2}{4}
   \left(\frac{k^+ \pa_t k^- + k^- \pa_t k^+}{k^-} +
     \frac{k^- \pa_t k^+ - k^+ \pa_t k^-}{k^-}
\,\frac{\sin^2\chi}{\chi^2}
   \right)
\eeqa
where $\chi = \sqrt{k^+ k^-}$. The energy density for the kaon-nucleon
sector
is given by the Hamiltonian
${\mathcal H}_{KN} = p_+ \pa_t k^+ - p_- \pa_t k^-  - {\mathcal
L}_{\chi}$
and it reads
\beqa
{\mathcal H}_{KN} &=&
\frac{f^2 {\left( k^+ \pa_t k^- + k^- \pa_t k^+ \right) }^2}{8
{\chi}^2} -
 \frac{f^2 {\left( k^+ \pa_t k^- - k^- \pa_t k^+ \right)
}^2\sin^2\chi}{8
{\chi }^4}  \\
& & + \left(2 f^2 m_K^2 + \left(2 a_2 + 4 a_3 \right)  m_s \,\bar{n} n +
    \left(2 a_1 + 2 a_2 + 4 a_3 \right) m_s \,\bar{p}p
\right)\sin^2\frac{\chi}{2} \nonumber
\eeqa
The charge density $n_K = i (p_+ k^+ - p_- k^-)$ becomes
\beq
n_K  =  \left( \bar{n}n + 2 \bar{p}p \right)\sin^2\frac{\chi}{2} +
   \frac{i f^2}{2} ( k^+ \pa_\mu k^- - k^- \pa_\mu k^+ )
 \frac{{\sin^2{\chi}}}{\chi^2}
\eeq
The Baym theorem for $k^\pm$ fields allows us to write them as $k^\pm =
\th
\exp^{\pm i \mu_K t}$
and with this form we get the equation
(\ref{endens-KN},\ref{kaon-dens}).

\subsection{Minimum of energy}
\label{app-a2}
The derivative of $\ep_{KN}$with respect to $\th$, keeping $n_i = n,x,n_K$
fixed, is
\beq
\left(\frac{\pa \ep_{KN}}{\pa \th} \right)_{n_i} =
\frac{\pa \ep_{KN}(n,x,\mu_K,\th)}{\pa \th} +
\frac{\pa \ep_{KN}(n,x,\mu_K,\th)}{\pa \mu_K}
\left(\frac{\pa\mu_K}{\pa\th} \right)_{n_i} . \nonumber
\label{der-en}
\eeq
It remains to find $\left(\fracd{\pa\mu_K}{\pa\th}\right)_{n_i}$.
This may be done by taking derivative of both sides of (\ref{kaon-dens})
\[
0 = f^2  \sin^2\th \left(\frac{\pa \mu_K}{\pa \th}\right)_{n_i}+ \;
   2 f^2 \mu_K \sin\th\cos\th \; + \; \fract{1}{2} n(1+x)\sin\th~.
\]
This approach is, of course, equivalent to minimization of a new
thermodynamic potential $\tilde{\ep} \equiv \ep_{KN} - \mu n_K$ 
with respect to $\th$ with $n,x,\mu$  fixed, as was done
in \cite{Thorsson:1994bu}.

\end{document}